\documentclass[]{interact}

\usepackage{epstopdf}
\usepackage[caption=false]{subfig}

\usepackage{listings} 

\usepackage[numbers,sort&compress]{natbib}
\bibpunct[, ]{[}{]}{,}{n}{,}{,}
\renewcommand\bibfont{\fontsize{10}{12}\selectfont}

\usepackage{xcolor} 

\theoremstyle{plain}

\theoremstyle{definition}

\theoremstyle{remark}

\begin{document}

\articletype{Original Research Article}

\title{Mistaken identities lead to missed opportunities: Testing for mean differences in partially matched data}

\author{
\name{Raymond Pomponio\textsuperscript{a}\thanks{CORRESPONDING AUTHOR: Raymond Pomponio. EMAIL: raymond.pomponio@cuanschutz.edu}, Bailey K. Fosdick\textsuperscript{a}, Julia Wrobel\textsuperscript{b} and Ryan A. Peterson\textsuperscript{a}}
\affil{\textsuperscript{a}Department of Biostatistics and Informatics, Colorado School of Public Health, University of Colorado-Denver Anschutz Medical Campus, Aurora, CO; \textsuperscript{b}Department of Biostatistics and Bioinformatics, Rollins School of Public Health, Emory University, Atlanta, GA}
}

\maketitle

\begin{abstract}
It is increasingly common to collect pre-post data with pseudonyms or self-constructed identifiers. On survey responses from sensitive populations, identifiers may be made optional to encourage higher response rates. The ability to match responses between pre- and post-intervention phases for every participant may be impossible in such applications, leaving practitioners with a choice between the paired $t$-test on the matched samples and the two-sample $t$-test on all samples for evaluating mean differences. We demonstrate the inadequacies with both approaches, as the former test requires discarding unmatched data, while the latter test ignores correlation and assumes independence. In cases with a subset of matched samples, an opportunity to achieve limited inference about the correlation exists. We propose a novel technique for such `partially matched' data, which we refer to as the Quantile-based $t$-test for correlated samples, to assess mean differences using a conservative estimate of the correlation between responses based on the matched subset. Critically, our approach does not discard unmatched samples, nor does it assume independence. Our results demonstrate that the proposed method yields nominal Type I error probability while affording more power than existing approaches. Practitioners can readily adopt our approach with basic statistical programming software.
\end{abstract}


\section{Introduction}

\begin{figure}
\centering
\subfloat[Matched data.]{
\resizebox*{4.5cm}{!}{\includegraphics{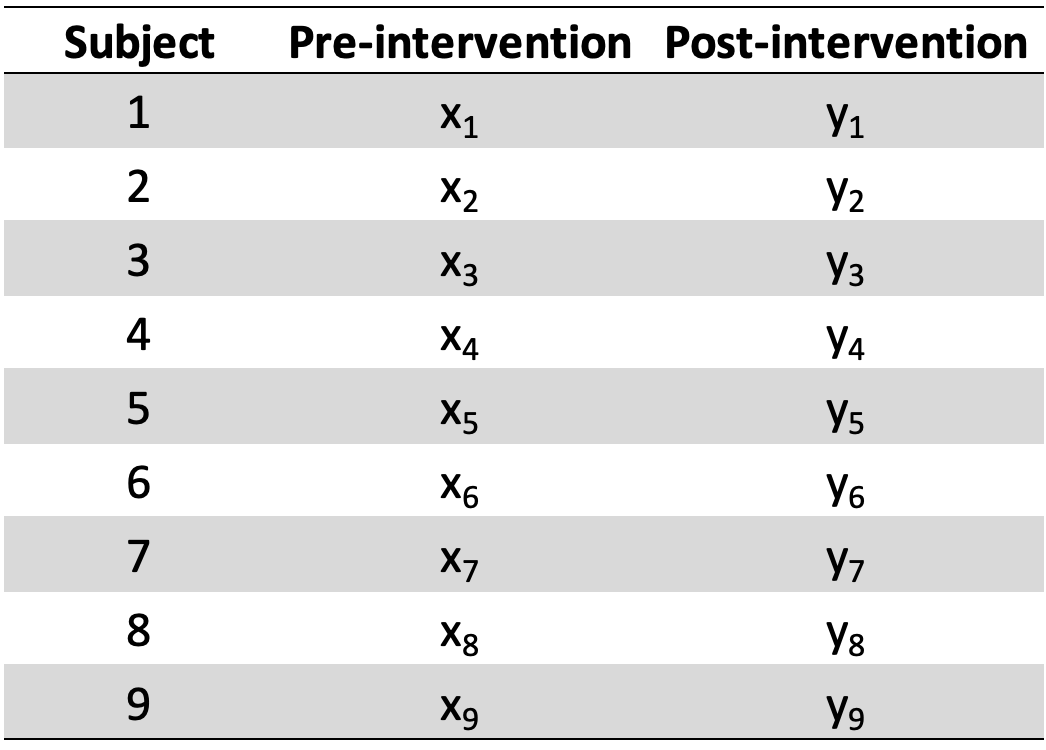}}}
\subfloat[Partially matched data.]{
\resizebox*{5.0cm}{!}{\includegraphics{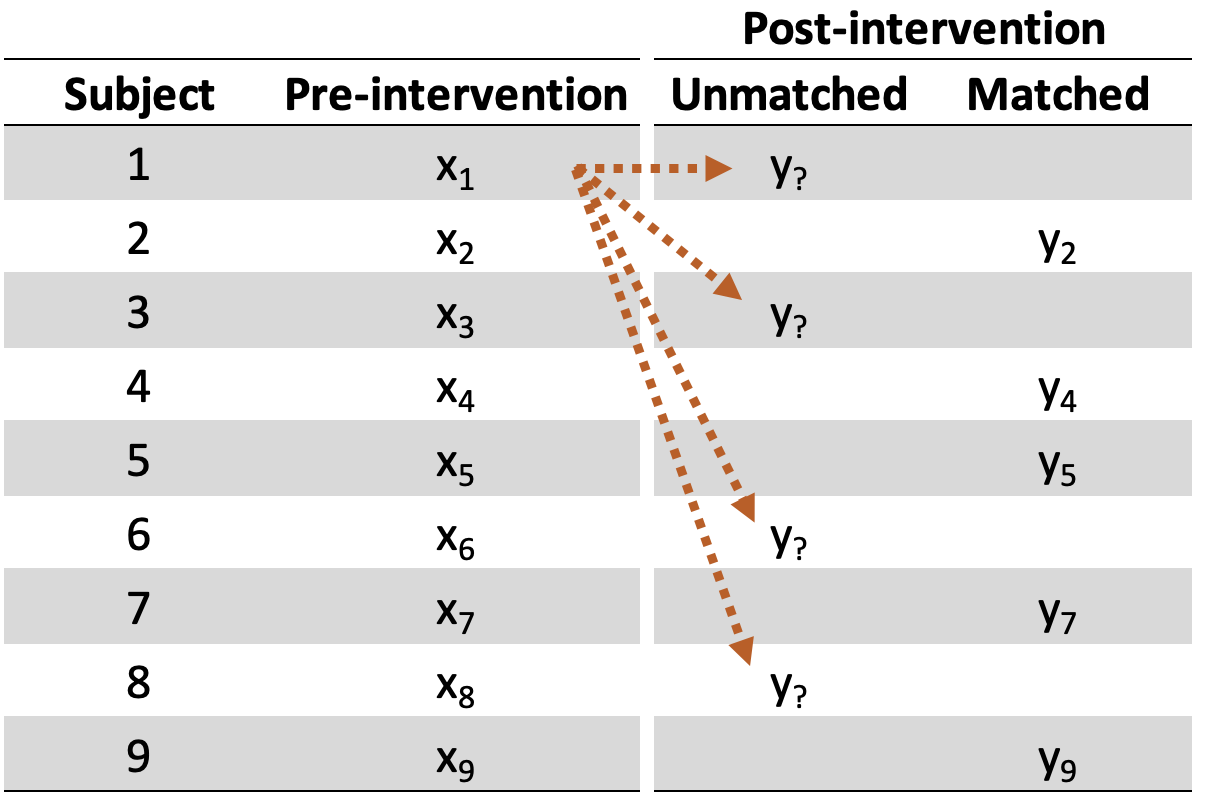}}}
\subfloat[Unmatched data.]{
\resizebox*{4.5cm}{!}{\includegraphics{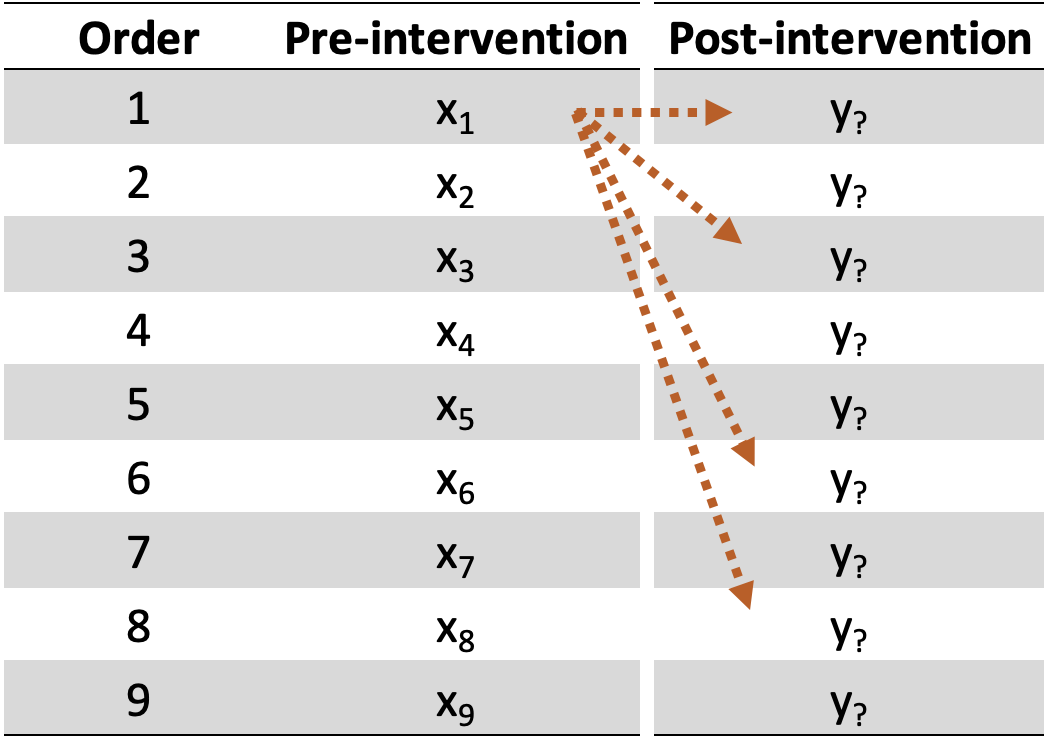}}}
\caption{Diagram illustrating varying matched data scenarios.} \label{fig:illustration}
\end{figure}

A number of intervention studies have examined pre-post differences in data that contain fully or partially anonymized identifiers \cite{McGloin96, Unwin22, McCauley23, Dunton14, Baird18, Caldwell2019}. For example, in \cite{Caldwell2019}, pediatric residents were anonymously surveyed before and after receiving a training on providing parenting advice. In \cite{McCauley23}, which surveyed high school students about vaping habits and knowledge, students were identified by the first three letters of one's maternal name, one's birth month, and the first three letters of the name of one's favorite teacher.

With such data, participants' identities are protected, but it may be impossible to match responses between pre-intervention and post-intervention. In pre-post studies with continuous outcomes, the paired $t$-test is the first choice for mean difference testing, however to use this approach practitioners are forced to discard unmatched responses leading to a loss of information.

Since anonymity is desirable in pre-post studies collecting sensitive information, researchers may ask participants to construct an identifier based on self-determined characteristics. However, such identification schemes are imperfect in that they rely on the consistency of each participant's answers to the same survey questions; the aforementioned study in \cite{McCauley23} was able to conclusively link responses from fewer than 60\% of participants in a sample of 600 students. Further, some participants may be uncomfortable providing such identifying information and may choose to self-anonymize by entering inconsistent information, or neglecting to answer questions entirely.

Recently, an approach was developed for testing mean differences in data that are entirely unmatched, or so called `unordered samples' data, which is based on sample splitting and sub-sampling \cite{Wang22}. Such data exist due to limitations in study design or concerns about confidentiality. The same issues may exist in partially matched data, but that some of the identities are recovered for matching. In one study, researchers described an oversight in data collection that prohibited the matching of pre-post surveys \cite{Knecht19}; whereas the paired $t$-test would have been used, investigators opted to use separate one-sample $t$-tests on the pre- and post-intervention responses. This was to avoid the problematic assumption of independence required for the two-sample $t$-test.

In Figure \ref{fig:illustration}, we illustrate the characteristics of matched, partially matched, and unmatched data. Matched data are characterized by the ability to conclusively link every pre-intervention response with a corresponding post-intervention response. If matching is impossible for each datum, then we consider the data unmatched. If some responses are able to be matched (say, for example, using self-constructed identifiers, or due to an optional identifier field), the partially matched data scenario exists, which is the focus of this paper.

With partially matched data, practitioners may face the dilemma of two rudimentary tests, the paired $t$-test using only the matched samples and the two-sample $t$-test using all samples. The paired $t$-test accommodates correlated samples by pairing (i.e., grouping) pre-post responses and analyzing the difference scores, while the two-sample $t$-test assumes independence. Yet in partially matched data, both approaches entail loss of power relative to a paired $t$-test using all samples. The current shortcomings of available methods may lead practitioners to confusion between two-sample $t$-test and the paired $t$-test \cite{Xu17}.

\begin{figure}
\centering
\subfloat[Power of available methods.]{%
\resizebox*{7.15cm}{!}{\includegraphics{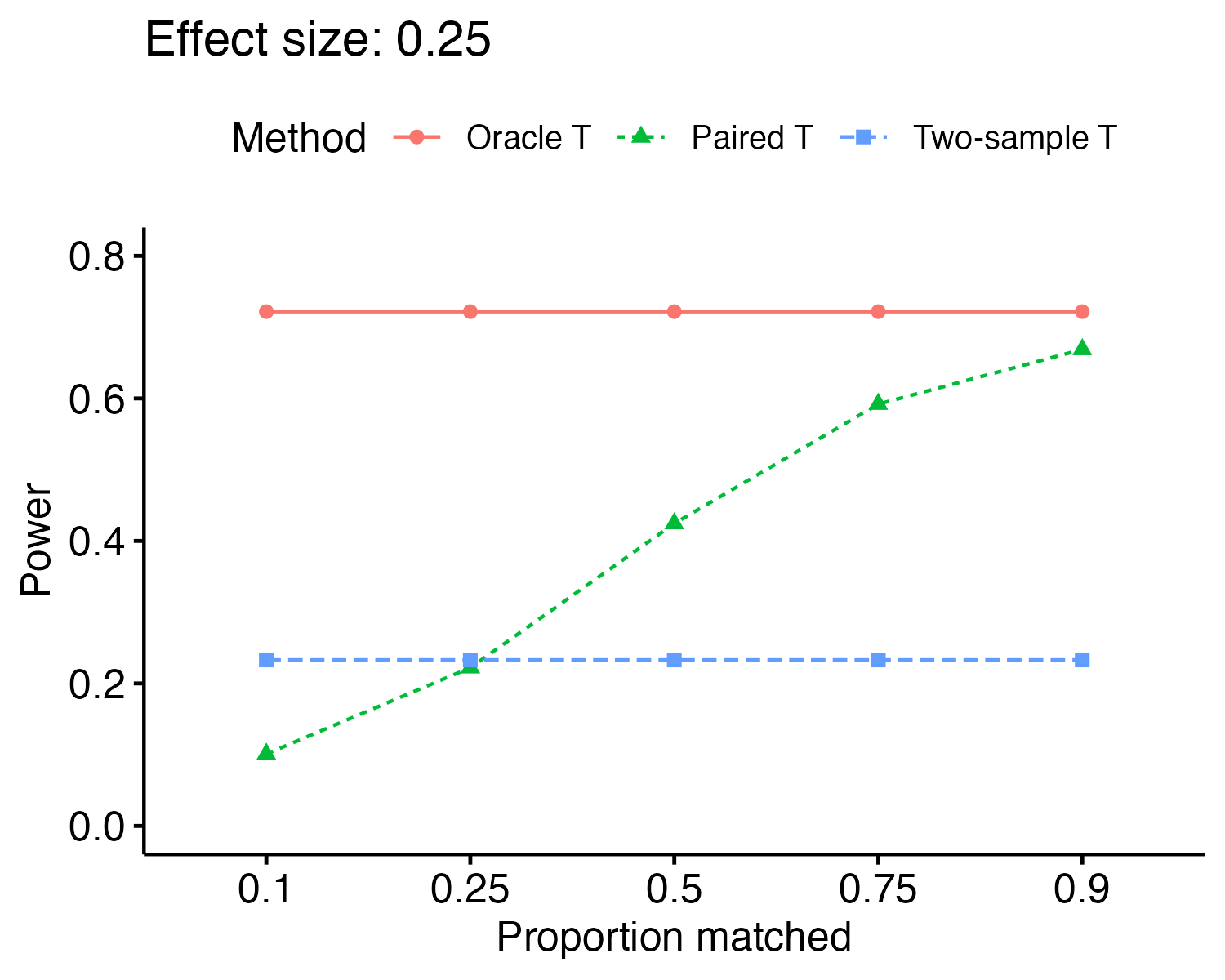}}}
\subfloat[Type I error probability of available methods.]{%
\resizebox*{7.15cm}{!}{\includegraphics{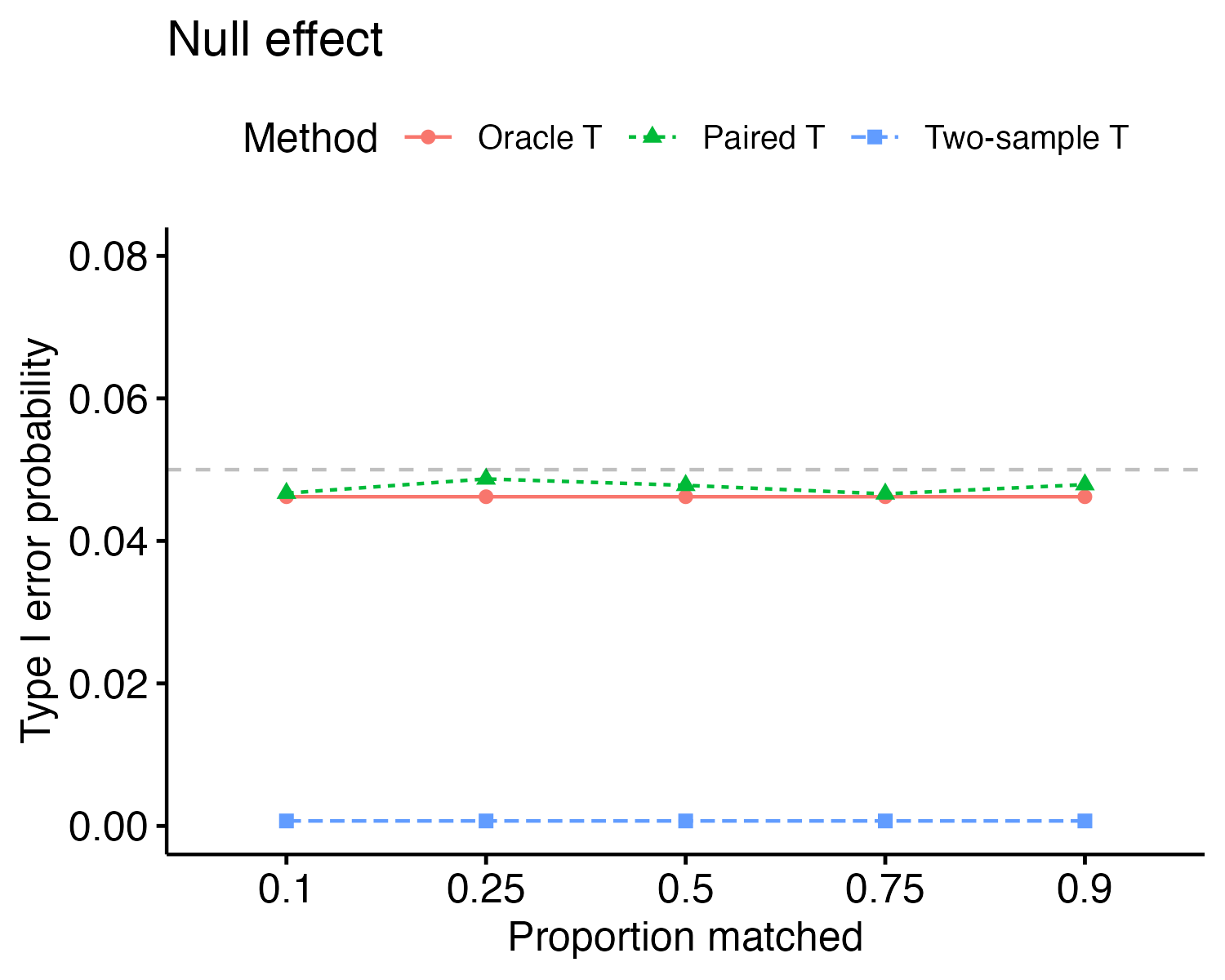}}}
\caption{The power and type I error of available methods, relative to the 'oracle' method (in red), for partially matched data. The paired $t$-test on matched samples is shown in green, and the two-sample $t$-test on all samples is shown in blue. All data were generated with a correlation of $0.65$ and a sample size of $75$. The effect of an increasing proportion of matched samples was examined.} \label{fig:power-gap}
\end{figure}

In Figure \ref{fig:power-gap}, we demonstrate the loss of power relative to an `oracle' $t$-test (the paired $t$-test using all samples, i.e., with all pairings known) exhibited by rudimentary methods (the paired $t$-test applied to the matched samples and the two-sample $t$-test applied to all samples). We simulated bivariate-normal data with sample size of $75$ and correlation of $0.65$. The approach employing the paired $t$-test reduces the sample size, and thus power, while the two-sample $t$-test is inefficient when samples are positively correlated. As expected, the paired $t$-test on matched samples approaches the same power as the oracle test when the proportion of matched samples approaches one. The two-sample $t$-test yields low Type I error probability compared to the other methods. Neither of the available tests are entirely suitable for partially matched data.

To our knowledge, partially matched data have received relatively little attention, despite an opportunity to estimate the correlation between samples using the matched subset and apply inference of the correlation to the entire sample of responses. An approach that properly addresses the correlated structure within pre-post intervention studies can improve power beyond the capabilities of existing methods for partially matched data.

This paper is organized as follows. In Section \ref{sec:methods}, we derive a novel approach for testing mean differences in partially matched data that uses a conservative estimate of the correlation between samples. In Section \ref{sec:results}, we demonstrate via simulation that our approach enables a more-powerful test while controlling Type I error probability, compared to existing alternatives for partially matched data. In Section \ref{sec:discussion}, we conclude with a discussion of possible complexities that may impede real-world analyses of partially matched data.

\section{Methods} \label{sec:methods}

\subsection{Notation}

Throughout the remainder of this paper we denote the vector of pre-intervention samples as $X$, and the vector of post-intervention samples as $Y$. We assume $x_i, y_i$ are paired samples from the bivariate normal distribution defined by the probability density

\small
\begin{multline}
        f(x_i, y_i)=\frac{1}{2\pi \sigma_X \sigma_Y \sqrt{1 - \rho^2}} \text{exp} \left[ \frac{(x_i - \mu_X)^2}{\sigma_X^2} - \frac{2 \rho (x_i - \mu_X) (y_i - \mu_Y)}{\sigma_X \sigma_Y} + \frac{(y_i - \mu_Y)^2}{\sigma_Y^2} \right]
\end{multline}
\normalsize

\noindent{where $\mu_X$ and $\mu_Y$ are respective pre- and post-intervention means, $\sigma_X^2$ and $\sigma_Y^2$ are respective pre- and post-intervention variances, and $\rho$ is the correlation between pre- and post-intervention measures.}

We use $n$ to denote the overall number of paired samples within a dataset. Lastly, we define the effect size as $\delta=\mu_X - \mu_Y$, which is the difference between means; it is the estimand of interest.

\subsection{Introducing the Quantile-based t-test for correlated samples} \label{sec:derivation}

The $t$-test for correlated samples \cite{Zimmerman93, Zimmerman12} is an alternative to the paired $t$-test that presumes known correlation, $\rho$. Under the assumption of equal variances ($\sigma_X=\sigma_Y$), the test statistic, which we denote $T'$, is computed by

\begin{equation} \label{eq:correlated_t}
    T'=\frac{\bar{X} - \bar{Y}}{\sqrt{\frac{s_X^2+s_Y^2}{n}(1-\rho})}
\end{equation}

\noindent{where $s_X^2$ and $s_Y^2$ are pre- and post-intervention sample variances, respectively.}

Under the null hypothesis, $T'$ is $t_{2n-2}$ distributed. Note, in the case of no correlation, $T'$ is identical to the two-sample $T$ statistic. However, correlation is not known in most practical circumstances and therefore one must use an estimate in place of $\rho$ in Equation (\ref{eq:correlated_t}). It has been shown that substituting the Pearson correlation for $\rho$ results in a test statistic that yields inflated Type I error probability in sample sizes up to $20$. In larger samples, i.e. $n=100$, the Type I error probability given the same approach is near-nominal \cite{Zimmerman12}.

The Type I error probability of the $t$-test for correlated samples is directly related to the standard error of the difference in means, which is influenced by $\rho$. Overestimating the correlation will yield an underestimate of the standard error and will inflate the Type I error probability of a test based on $T'$. We now introduce a conservative estimate of $\rho$ such that the standard error is appropriately estimated, yielding a new  statistic $T_{q^*}'$, with a Type I error probability that is controlled at or near the nominal level.

Our conservative estimate of $\rho$ is based on a quantile estimate, which we denote $r_q$ (the $20^{th}$ quantile estimate, for example is $r_{q20}$). The quantile estimate is derived from the lower bound of a one-sided confidence interval for the correlation between samples. A confidence interval for the correlation can be derived using Fisher's $z$ transformation and is detailed in Appendix \ref{app:derivation}. Note that the confidence interval requires at least four matched samples to be calculable, due to the term $n - 3$ in the denominator of the standard error formula.

Our approach relies on simulation to select a context-dependent quantile, which targets the desired nominal Type I error probability. We refer to this procedure as `$\alpha$-targeting', since the aim is to achieve a test with near-nominal Type I error probability, or $\alpha$. We simulated bivariate normal datasets with varying sample size, proportion of matched samples, and correlation. For each combination of simulation settings, we generated $10,000$ datasets. We then computed the Type I error probability of an array of tests using evenly-spaced quantiles $0.15, 0.20, ..., 0.50$ to estimate $\rho$ in Equation \ref{eq:correlated_t}. We identified the quantile that resulted in closest-to-nominal Type I error probability for each combination of simulation settings. For this analysis, we set the nominal Type I error probability to $\alpha=0.05$.

Once identified via simulation, the $\alpha$-targeted quantile is used in our quantile-based $T'$ statistic by computing

\begin{equation}
    T_{q^*}'=\frac{\bar{X} - \bar{Y}}{\sqrt{\frac{s_X^2+s_Y^2}{n}(1-r_{q^*}})}
\end{equation}

\noindent{where $q^*$ is the $\alpha$-targeted quantile of the correlation estimate. We report the resulting quantiles that were selected for various simulation contexts. Specifically, we display the 5\% $\alpha$-targeted quantile values for different sample sizes, matched sample proportions, and true correlations in the bivariate normal case, so that in practice, these quantiles can be utilized to target the 5\% nominal Type I error rate given a study-specific sample size and proportion matched.}

\subsection{Performance comparison of available methods}

We compared the performance of our quantile-based $T'$ statistic with available alternatives, namely, the two-sample $T$ statistic and the paired $T$ statistic applied to matched samples only (we refer to the latter method as the `matched, paired $T$'). Our performance measures were Type I error probability and power, where an optimal method would yield empirical Type I error probability close to nominal and the highest power. Type I error probability in this case was computed by the proportion of rejected null hypotheses when the effect size was zero. Power was computed by the proportion of rejected null hypotheses when the effect size was positive.

We simulated bivariate normal datasets with varying sample sizes and proportions of matched samples. For each combination of simulation settings, we generated $10,000$ datasets, each with a uniformly random correlation between $\rho=0.1$ and $\rho=0.9$. For simplicity, we set $\sigma_X=\sigma_Y=1$ and assumed equal variance between pre- and post-intervention responses for all methods. We used the smallest $\alpha$-targeted quantile for a given combination of sample size and matched sample proportion, based on the results of our simulation described in the previous section. This approach was conservative and mimicked real applications in which correlation is not known.

We compared the Type I error probability and power of the three competing methods, along with a fourth method that employed the Pearson correlation of matched samples (we refer to this method as the `Pearson-based $T'$'). This last approach was included to demonstrate the relative merits of a commonly-available estimate of the correlation versus our proposed quantile-based estimate.

We analyzed the performance of each method and the relationship between relative test performance and sample size, as well as matched-sample proportion. We also estimated the expected Type I error probability for varying correlations by method using logistic regression with the results of our simulation.

\section{Results} \label{sec:results}

\subsection{Alpha-targeted quantiles for the proposed T statistic}

Our proposed approach employs a quantile-based estimate of the correlation between samples using the matched subset for inference. The results of the $\alpha$-targeted quantile procedure for a range of data conditions are presented in Table \ref{tab:selected-quantiles}.

\begin{table}
\tbl{$\alpha$-targeted quantiles found for the Quantile-based $t$-test for correlated samples, with varying sample size, correlation, and proportion of matched samples.\textsuperscript{a, b}}
{\begin{tabular}{lcccccc} \toprule
 & & \multicolumn{5}{l}{Proportion of matched samples} \\ \cmidrule{3-7}
 N & Cor. ($\rho$) & 0.1 & 0.25 & 0.5 & 0.75 & 0.9 \\ \midrule
    20 & 0.1 & - & 0.25 & 0.35 & 0.35 & 0.4 \\
    20 & 0.25 & - & 0.25 & 0.35 & 0.35 & 0.4 \\
    20 & 0.5 & - & 0.25 & 0.3 & 0.35 & 0.4 \\
    20 & 0.9 & - & 0.2 & 0.3 & 0.35 & 0.4 \\
    \midrule
    50 & 0.1 & 0.25 & 0.35 & 0.35 & 0.4 & 0.4 \\
    50 & 0.25 & 0.25 & 0.35 & 0.35 & 0.4 & 0.4 \\
    50 & 0.5 & 0.25 & 0.35 & 0.35 & 0.4 & 0.4 \\
    50 & 0.9 & 0.2 & 0.3 & 0.35 & 0.4 & 0.4 \\
    \midrule
    100 & 0.1 & 0.3 & 0.35 & 0.4 & 0.4 & 0.4 \\
    100 & 0.25 & 0.3 & 0.35 & 0.4 & 0.4 & 0.4 \\
    100 & 0.5 & 0.3 & 0.35 & 0.4 & 0.4 & 0.4 \\
    100 & 0.9 & 0.25 & 0.35 & 0.4 & 0.4 & 0.45 \\
    \midrule
    200 & 0.1 & 0.35 & 0.4 & 0.35 & 0.4 & 0.4 \\
    200 & 0.25 & 0.35 & 0.4 & 0.4 & 0.4 & 0.4 \\
    200 & 0.5 & 0.35 & 0.4 & 0.4 & 0.4 & 0.4 \\
    200 & 0.9 & 0.35 & 0.4 & 0.4 & 0.4 & 0.45 \\ \bottomrule
\end{tabular}}
\tabnote{\textsuperscript{a}Data were simulated 10,000 times for each combination of settings.

\textsuperscript{b} For scenarios with fewer than four matched samples, the Quantile-based $T'$ statistic is not calculable.}
\label{tab:selected-quantiles}
\end{table}

$\alpha$-targeted quantiles were identified via simulation to yield near-nominal Type I error probability. For a fixed sample size and proportion of matched samples, correlation did not heavily influence the selected quantile. When correlation impacted the selected quantile, quantiles were generally within $0.05$ of one another for the same sample sample size and matched sample proportion. For example, at a sample size of $50$ with a matched sample proportion of $0.25$, the selected quantile was $0.35$ with correlations of ${0.1, 0.25, 0.5}$, but was $0.3$ when correlation was $0.9$. Thus, without knowing the correlation in advance, but anticipating a positive magnitude between $\rho=0.1$ and $\rho=0.9$, we can identify a suitable quantile for a given context to within $\pm 5$ percentage points. Since correlation is typically not known, a conservative approach is to use the minimum $\alpha$-targeted quantile for a given combination of sample size and matched sample proportion, since lower quantiles yield higher standard errors.

For a fixed sample size, the $\alpha$-targeted quantile increased as the proportion of matched samples increased. Thus, with more matched data, one should use a larger quantile, which more closely approximates the Pearson correlation of the matched samples. Similarly, for a fixed proportion of matched samples, the $\alpha$-targeted quantile increased as the sample size increased. With a matched sample proportion of $0.25$ and a sample size of $20$, one should use a quantile of $0.2$ to yield a test with near-nominal Type I error probability. However, for the same proportion of matched samples at a sample size of $200$, one should use a quantile of $0.4$ for nominal Type I error probability.

In the next section, we compared the simulated performance of available methods to our approach, which employs the $\alpha$-targeted quantile as an estimate of the correlation along with the $t$-test for correlated samples.

\subsection{Simulated performance comparison of available methods}

We report the results of our performance comparison in Table \ref{tab:performance-comp}. For a variety of partially matched data conditions, we report the null hypothesis rejection rates for effect sizes of $\delta=0$, $\delta=0.25$, and $\delta=0.5$, respectively representing a null effect, a medium-strength effect, and a strong effect.

In the null scenario, the two-sample $t$-test generally yielded sub-nominal Type I error probability. Note the Type I error probability of the two-sample $t$-test was unaffected by the matched sample proportion since the test assumes independence and uses the full dataset regardless of the proportion of matched samples. In contrast, the paired $t$-test applied to matched samples yielded nominal Type I error probability. Our proposed Quantile-based $t$-test for correlated samples generally yielded nominal Type I error probability but occasionally erred on the side of sub-nominal Type I error probability. The latter result is a reflection that our quantile-based method tended to be conservative in Type I error. Importantly, the Pearson-based $t$-test for correlated samples tended to yield inflated Type I error probability, especially with smaller sample sizes and with lower proportions of matched samples.

In the medium-strength effect scenario, our proposed method tended to afford greater power than the two-sample $t$-test and the paired $t$-test applied to matched samples. For example, in samples of size $50$ with matched sample proportions of $0.5$, the Quantile-based $t$-test for correlated samples yielded power of $0.462$, versus $0.151$ for that of the two-sample $t$-test, and $0.289$ for that of the paired $t$-test applied to matched samples. The difference in power between our proposed test and the paired $t$-test generally declined as the sample size increased and the propotion of matched samples increased, though in scenarios with matched sample proportions of $0.1$, our method consistently outperformed the paired $t$-test.

Results were generally similar for the strong effect scenario, with two differences. First, the two-sample $t$-test was more powerful than the paired $t$-test applied to matched samples in the case of $\delta=0.5$, though it was about equally as powerful as our proposed test. Thus, when the effect size was large, ignoring correlation altogether was more powerful than analyzing only matched samples. Second, all methods tended to afford power near one when samples were large enough, generally at or above $n=100$, and when the proportion of matched samples was high, generally $0.5$ or greater. The difference between methods in terms of power was therefore less substantial in larger samples with large effects. We expect that the results with large sample sizes presented here approximate the asymptotic properties of the methods explored.

The Pearson-based $t$-test for correlated samples was typically more powerful than the corresponding Quantile-based $t$-test, though it tended to inflate the Type I error probability. The power gains achievable with the Pearson-based method were generally greatest in smaller samples. For example, in the medium-strength effect scenario, at a sample size of $20$ with matched sample proportions of $0.5$, the Quantile-based $t$-test for correlated samples yielded power of $0.203$, versus $0.282$ for that of the Pearson-based $t$-test for correlated samples.

For conditions with fewer than four matched samples, the Quantile-based $t$-test for correlated samples was not estimable due to the requirement of four or more matched samples to compute the correlation estimate, previously mentioned in Section \ref{sec:derivation}. However, this only occurred in scenarios with a sample size of $20$ of which 10\% were matched. Similarly, results were withheld for the Pearson-based test in these scenarios, due to the restricted support of the Pearson correlation with only two samples; in such cases it is only estimable as $-1$, $0$, or $1$, which led to inconsistent results in Type I error probability.

\begin{sidewaystable}
    \caption{Performance comparison of available methods based on simulated Type I error probabilities and Power for varying sample size and proportion of matched samples, using 10,000 simulations for each setting.}
    \centering
    \begin{tabular}{llccccccccc} \toprule
     & & \multicolumn{9}{c}{Effect size \textsuperscript{a}} \\
     & & \multicolumn{3}{|c|}{$\delta=0$} & \multicolumn{3}{|c|}{$\delta=0.25$} & \multicolumn{3}{|c|}{$\delta=0.5$} \\ \cmidrule{3-11}
    & & \multicolumn{9}{c}{Matched sample proportion} \\
    N & Method & 0.1 & 0.5 & 0.9 & 0.1 & 0.5 & 0.9 & 0.1 & 0.5 & 0.9 \\ \midrule
    20 & Two-sample $T$ & \textcolor{red}{0.014} & \textcolor{red}{0.014} & \textcolor{red}{0.014} & 0.057 & 0.057 & 0.057 & 0.288 & 0.288 & 0.288 \\
    20 & Matched, Paired $T$ & 0.053 & 0.051 & 0.053 & 0.056 & 0.137 & 0.218 & 0.067 & 0.360 & 0.575 \\
    20 & Quantile-based $T'$ \textsuperscript{b} & - & 0.047 & 0.053 & - & 0.203 & 0.242 & - & 0.564 & 0.618 \\
    20 & Pearson-based $T'$ \textsuperscript{c} & - & \textcolor{red}{0.082} & \textcolor{red}{0.066} & - & 0.282 & 0.271 & - & 0.643 & 0.646 \\ \midrule
    50 & Two-sample $T$ & \textcolor{red}{0.012} & \textcolor{red}{0.012} & \textcolor{red}{0.012} & 0.151 & 0.151 & 0.151 & 0.778 & 0.778 & 0.778 \\
    50 & Matched, Paired $T$ & 0.046 & 0.051 & 0.052 & 0.080 & 0.289 & 0.453 & 0.176 & 0.701 & 0.882 \\
    50 & Quantile-based $T'$ & \textcolor{red}{0.041} & 0.045 & 0.047 & 0.297 & 0.462 & 0.477 & 0.776 & 0.892 & 0.900 \\
    50 & Pearson-based $T'$ & \textcolor{red}{0.131} & \textcolor{red}{0.059} & 0.054 & 0.512 & 0.502 & 0.497 & 0.886 & 0.907 & 0.908 \\ \midrule
    100 & Two-sample $T$ & \textcolor{red}{0.012} & \textcolor{red}{0.012} & \textcolor{red}{0.012} & 0.382 & 0.382 & 0.382 & 0.981 & 0.981 & 0.981 \\
    100 & Matched, Paired $T$ & 0.054 & 0.050 & 0.050 & 0.133 & 0.481 & 0.682 & 0.359 & 0.905 & 0.986 \\
    100 & Quantile-based $T'$ & \textcolor{red}{0.041} & 0.048 & 0.047 & 0.618 & 0.708 & 0.712 & 0.980 & 0.991 & 0.991 \\
    100 & Pearson-based $T'$ & \textcolor{red}{0.083} & 0.056 & 0.052 & 0.712 & 0.723 & 0.722 & 0.989 & 0.991 & 0.992 \\ \midrule
    200 & Two-sample $T$ & \textcolor{red}{0.009} & \textcolor{red}{0.009} & \textcolor{red}{0.009} & 0.791 & 0.791 & 0.791 & 1.000 & 1.000 & 1.000 \\
    200 & Matched, Paired $T$ & 0.047 & 0.052 & 0.052 & 0.236 & 0.723 & 0.890 & 0.612 & 0.992 & 1.000 \\
    200 & Quantile-based $T'$ & 0.050 & 0.046 & 0.048 & 0.891 & 0.907 & 0.910 & 1.000 & 1.000 & 1.000 \\ 
    200 & Pearson-based $T'$ & \textcolor{red}{0.066} & 0.053 & 0.053 & 0.909 & 0.913 & 0.914 & 1.000 & 1.000 & 1.000 \\ \bottomrule
    \end{tabular}
    \tabnote{\textsuperscript{a} Note an effect size of $\delta=0$ represents the null scenario and the corresponding rejection rates are Type I error probabilities. Results that fall outside 3 Monte Carlo standard errors are highlighted in \textcolor{red}{red}.
    \textsuperscript{b} For scenarios with fewer than four matched samples, the Quantile-based $T'$ statistic is not calculable.
    \textsuperscript{c} For scenarios with two matched samples, the Pearson correlation is only calculable as $-1$, $0$, or $1$. Results were withheld due to inconsistent rejection rates for these scenarios.}
    \label{tab:performance-comp}
    
\end{sidewaystable}

\subsection{Expected Type I error probability versus correlation}

\begin{figure}
\centering
\includegraphics[width=13cm]{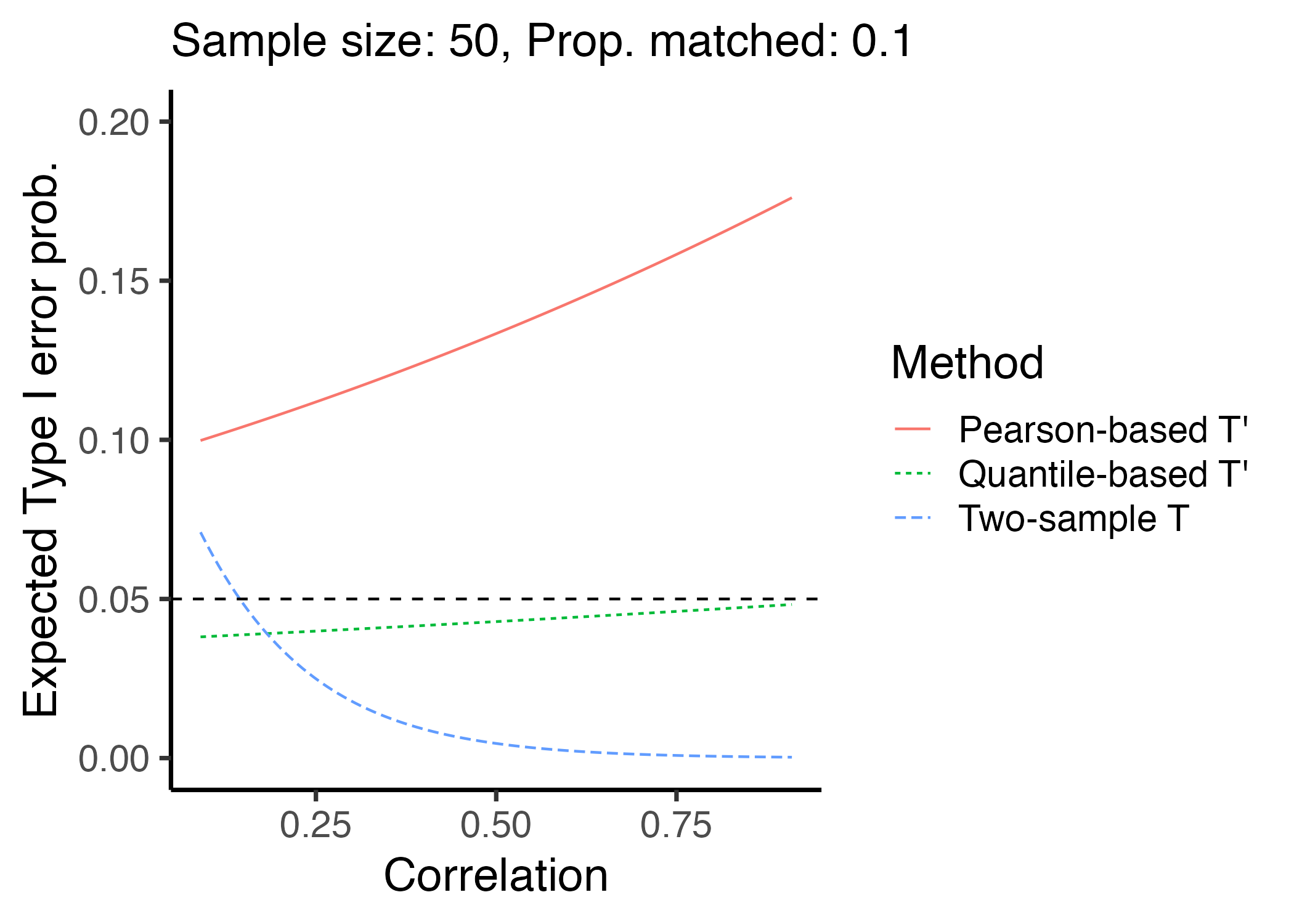}
\caption{The expected Type I error probability as a function of correlation, for a sample size of $50$ with a matched sample proportion of $0.1$, based on simulation results. The proposed method, the Quantile-based $t$-test for correlated samples, is plotted in green; the Pearson-based $t$-test for correlated samples is plotted in red; the two-sample $t$-test is plotted in blue. The nominal Type I error probability is dotted in black. Curves were estimates using logistic regression on 10,000 simulations.} \label{fig:reject-rate}
\end{figure}

Using the results of our null scenario simulation presented in the previous section, we analyzed the relationship between Type I error probability and correlation, using logistic regression to estimate the expectation at each correlation between $0.1$ and $0.9$. Figure \ref{fig:reject-rate} presents the resulting curves for a scenario with a samples size of $50$, of which 10\% were matched. The red line is the expected Type I error probability of the Pearson-based $t$-test for correlated samples. The blue line is the expected Type I error probability of the two-sample $t$-test. The green line is the expected Type I error probability of the Quantile-based $t$-test for correlated samples. The dotted line is included for visual aid; it represents the nominal Type I error probability.

It is apparent in this scenario that the Pearson-based $t$-test for correlated samples resulted in inflated Type I error probability, on average, for all positive correlations. This approach exhibited increasing Type I error inflation with higher correlations. The Quantile-based $t$-test achieved near-nominal Type I error probability over the domain, evidenced by the flatter curve near the dotted line. The two-sample $t$-test exhibited near-nominal Type I error probability when correlation was near-zero; however, at larger correlations the two-sample $t$-test exhibited sub-nominal Type I error probability.

\section{Discussion} \label{sec:discussion}

We evaluated a novel approach for testing mean differences in partially matched data that is well-powered and controls Type I error probability at the nominal rate. We demonstrated these properties using extensive simulations with 10,000 repetitions in each combination of settings. Thus, we expect our results to generalize to similar data; namely, data that are bivariate-normally distributed.

There are limitations to our analysis that are worth considering in real-world applications. One is that we have focused on normally distributed outcomes. In many applications, categorical and nominal outcomes are common. For example, an Likert-style scale from $1$ to $5$ may be used \cite{Likert32}, as was seen in \cite{McCauley23}. For such data, normally distributed outcomes cannot be assumed. Our approach might not be appropriate for data with outcomes involving relatively few unique values. However, in this limitation our method is not alone; the two-sample $t$-test and the paired $t$-test suffer the same issues. In guiding the decision of which hypothesis test practitioners should apply, we advise a look at the empirical distributions of the outcomes, considering transformations if necessary (for example, as in \cite{Zimmerman12}, a rank transformation was applied to non-normally distributed outcomes).

In addition, we have focused primarily on partially matched data that are deterministically linked. That is, no uncertainty exists in the linkage of matched pairs. This is accomplished via a matching strategy that links unit level identifiers. Recently others have elucidated methods for paired data that are probabilistically linked \cite{Sayers15}. We cannot claim our method is suitable for such data, although the record linkage problem may be considered a distinct challenge from the mean inference problem. Whereas we have focused on the latter, there is potential to develop an approach that unifies both the record linkage and mean inference procedures. We leave this possibility for future work. 

The challenges of partially matched data are not the same as those of so-called `partially paired' data \cite{Guo21, Guo17}. Such data have been explored under a variety of names (including `partially overlapping' and `partially correlated' data) in several recent papers \cite{Derrick22, Qin18, Derrick17, Samawi14, Dahiya80}. In such data, samples that are not paired are assumed to be independent. In contrast, a partially matched structure assumes unmatched samples to be correlated (we commonly anticipate positive correlation). However, real-world data may present a hybrid of these two scenarios. An example would be a pre-post study where pseudonyms are collected as identifiers, and where participants are lost to follow-up in the one of the data collection phases. We emphasize the challenge in analyzing such data, which may present identifiability concerns, as in determining whether a particular response is unmatched or unpaired. In future work, we plan to introduce an estimation approach that can be applied to this kind of data under certain limiting assumptions (i.e., assuming dropout occurs only at one point in time). While we believe our current work is pertinent to many existing applications, we recognize the extended utility of an approach that accommodates data that are both partially matched and partially paired.

We have focused our analysis through the simplifying assumption that variances in pre- and post- data are equal. This may not be realistic for all applications, and caution is warranted where variance in the pre-intervention phase may differ substantially from that of the post-intervention phase. Similar to the above discussion of non-normally distributed outcomes, we advise practitioners to examine their data distributions and consider transformations if necessary. If practitioners would like to run their own simulations to select optimal quantiles for unique circumstances, we provide code for doing so in Appendix \ref{app:simulation}. 

While methods exist for testing mean differences in unmatched data \cite{Wang22, Spruill82}, this paper has focused on paired data that contain a subset of matched samples. We used a conservative estimate of the correlation between samples to test for mean differences. The selection of a quantile is required, but we provide simulation-based selected quantiles for a variety of scenarios using a common alpha ($\alpha=0.05$), which practitioners can readily reference for future applications. We showed that our Quantile-based $t$-test for correlated samples yields nominal Type I error probability, while the corresponding Pearson-based test yields inflated Type I error probability in small to modestly-sized samples. Finally, we demonstrated that our approach is consistently as-or-more powerful than the paired $t$-test applied to matched samples only, as well as the two-sample $t$-test applied to all samples.

\section*{Acknowledgement(s)}

This work utilized the Alpine high performance computing resource at the University of Colorado Boulder. Alpine is jointly funded by the University of Colorado Boulder, the University of Colorado Anschutz, Colorado State University, and the National Science Foundation (award 2201538).

\section*{Disclosure statement}

The authors report there are no competing interests to declare.

\bigskip

\bibliographystyle{tfs}
\bibliography{interacttfssample}

\begin{thebibliography}{10}
\providecommand{\MR}{\relax\unskip\space MR }
\providecommand{\url}[1]{\normalfont{#1}}
\providecommand{\urlprefix}{Available at }

\bibitem{Baird18}
K. Baird, D.K. Creedy, A.S. Saito, and J. Eustace, \emph{Longitudinal
  evaluation of a training program to promote routine antenatal enquiry for
  domestic violence by midwives}, Women and Birth 31 (2018), pp. 398--406.
  \urlprefix\url{https://www.sciencedirect.com/science/article/pii/S1871519217305887}.

\bibitem{Caldwell2019}
A. Caldwell, H. Qasimyar, L. Shumate, M.P. Anderson, A. Cherry, C. Bryant, and
  A. Bax, \emph{Structured curriculum to improve pediatric resident confidence
  and skills in providing parenting advice}, Journal of Osteopathic Medicine
  119 (2019), pp. 748--755.
  \urlprefix\url{https://doi.org/10.7556/jaoa.2019.124}.

\bibitem{Dahiya80}
R.C. Dahiya and R.M. Korwar, \emph{{Maximum Likelihood Estimates for a
  Bivariate Normal Distribution with Missing Data}}, The Annals of Statistics 8
  (1980), pp. 687 -- 692.
  \urlprefix\url{https://doi.org/10.1214/aos/1176345020}.

\bibitem{Derrick17}
B. Derrick, B. Russ, D. Toher, and P. White, \emph{Test statistics for the
  comparison of means for two samples that include both paired and independent
  observations}, Journal of Modern Applied Statistical Methods 16 (2017), pp.
  137--157.

\bibitem{Derrick22}
B. Derrick and P. White, \emph{Review of the partially overlapping samples
  framework: Paired observations and independent observations in two samples},
  The Quantitative Methods for Psychology 18 (2022), pp. 55--65.
  \urlprefix\url{http://www.tqmp.org/RegularArticles/vol18-1/p055/p055.pdf}.

\bibitem{Dunton14}
G.F. Dunton, Y. Liao, R. Grana, R. Lagloire, N. Riggs, C.P. Chou, and T.
  Robertson, \emph{State-wide dissemination of a school-based nutrition
  education programme: a re-aim (reach, efficacy, adoption, implementation,
  maintenance) analysis}, Public Health Nutrition 17 (2014), p. 422–430.

\bibitem{Fisher1921}
R. Fisher, \emph{On the 'probable error' of a coefficient of correlation
  deduced from a small sample.}, Metron 1 (1921), pp. 3--32.

\bibitem{Guo17}
B. Guo and Y. Yuan, \emph{A comparative review of methods for comparing means
  using partially paired data}, Statistical Methods in Medical Research 26
  (2017), pp. 1323--1340.
  \urlprefix\url{https://doi.org/10.1177/0962280215577111}, PMID: 25834090.

\bibitem{Guo21}
X. Guo, Y. Wang, N. Zhou, and X. Zhu, \emph{Optimal weighted two-sample t-test
  with partially paired data in a unified framework}, Journal of Applied
  Statistics 48 (2021), pp. 961--976.
  \urlprefix\url{https://doi.org/10.1080/02664763.2020.1753027}.

\bibitem{Knecht19}
L.D. Knecht, K.J. Wilson, M.E. Linton, J.M. Koonmen, and E.F. Johns,
  \emph{Assessing student expectations and perceptions of a short-term
  international service-learning experience}, Public Health Nursing 37 (2020),
  pp. 121--129.
  \urlprefix\url{https://onlinelibrary.wiley.com/doi/abs/10.1111/phn.12669}.

\bibitem{Likert32}
R. Likertt, \emph{A technique for the measurement of attitudes}, Archives of
  Psychology 22 (1932), pp. 140--155.

\bibitem{McCauley23}
D.M. McCauley, M. Baiocchi, S. Cruse, and B. Halpern-Felsher, \emph{Effects of
  a short school-based vaping prevention program for high school students},
  Preventive Medicine Reports 33 (2023), p. 102184.
  \urlprefix\url{https://www.sciencedirect.com/science/article/pii/S221133552300075X}.

\bibitem{McGloin96}
J. McGloin, S. Holcomb, and D.S. Main, \emph{Matching anonymous pre-posttests
  using subject-generated information}, Evaluation Review 20 (1996), pp.
  724--736.

\bibitem{Qin18}
H. Qin, E. Prentice, and K. Freeman, \emph{Analyzing partially correlated
  longitudinal data in community survey research}, Society \& Natural Resources
  31 (2018), pp. 142--149.
  \urlprefix\url{https://doi.org/10.1080/08941920.2016.1264650}.

\bibitem{Samawi14}
H.M. Samawi and R. Vogel, \emph{Notes on two sample tests for partially
  correlated (paired) data}, Journal of Applied Statistics 41 (2014), pp.
  109--117. \urlprefix\url{https://doi.org/10.1080/02664763.2013.830285}.

\bibitem{Sayers15}
A. Sayers, Y. Ben-Shlomo, A.W. Blom, and F. Steele, \emph{{Probabilistic record
  linkage}}, International Journal of Epidemiology 45 (2015), pp. 954--964.
  \urlprefix\url{https://doi.org/10.1093/ije/dyv322}.

\bibitem{Spruill82}
N.L. Spruill and J.L. Gastwirth, \emph{On the estimation of the correlation
  coefficient from grouped data}, Journal of the American Statistical
  Association 77 (1982), pp. 614--620.
  \urlprefix\url{http://www.jstor.org/stable/2287724}.

\bibitem{Unwin22}
J. Unwin, C. Delon, H. Giæver, C. Kennedy, M. Painschab, F. Sandin, C.S.
  Poulsen, and D.A. Wiss, \emph{Low carbohydrate and psychoeducational programs
  show promise for the treatment of ultra-processed food addiction}, Frontiers
  in Psychiatry 13 (2022).
  \urlprefix\url{https://www.frontiersin.org/articles/10.3389/fpsyt.2022.1005523}.

\bibitem{Wang22}
Y. Wang, Y. Tang, and Z.S. Ye, \emph{{Paired or Partially Paired Two-sample
  Tests With Unordered Samples}}, Journal of the Royal Statistical Society
  Series B: Statistical Methodology 84 (2022), pp. 1503--1525.
  \urlprefix\url{https://doi.org/10.1111/rssb.12541}.

\bibitem{Xu17}
M. Xu, D. Fralick, J.Z. Zheng, B. Wang, X.M. Tu, and C. Feng, \emph{The
  differences and similarities between two-sample t-test and paired t-test.},
  Shanghai archives of psychiatry 29 (2017), pp. 184--188.

\bibitem{Zimmerman12}
D.W. Zimmerman, \emph{Correcting two-sample z and t tests for correlation: An
  alternative to one-sample tests on difference scores}, Psicológica 33
  (2012), pp. 391--418.

\bibitem{Zimmerman93}
D.W. Zimmerman, R.H. Williams, and B.D. Zurabo, \emph{Effect of nonindependence
  of sample observations on some parametric and nonparametric statistical
  tests}, Communications in Statistics - Simulation and Computation 22 (1993),
  pp. 779--789. \urlprefix\url{https://doi.org/10.1080/03610919308813123}.

\end{thebibliography}

\bigskip



\newpage

\appendix

\section{Derivation of the quantile correlation estimate} \label{app:derivation}

For the following derivation we denote the Pearson correlation estimate as $r$. A confidence interval for $r$ can be derived using Fisher's $z$ transformation \cite{Fisher1921},

\begin{equation} \label{eq:z_transformation}
    z=\text{arctanh}(r)=\frac{1}{2}\text{ln} \left(\frac{1 + r}{1 - r} \right)
\end{equation}

\noindent{where the quantity $z$ has an approximate normal distribution with standard error $SE_z=1/\sqrt{n - 3}$. Inverting the transformation will yield an approximate confidence interval for the correlation,}

\begin{equation} \label{eq:inverted_z}
    CI_{1-\alpha}(\rho)=
    \left\{\text{tanh}[\text{arctanh}(r - z_{\alpha/2} \times SE_z)],
    \text{tanh}[\text{arctanh}(r + z_{\alpha/2} \times SE_z)] \right\}
\end{equation}

\noindent{where $z_{\alpha/2}$ is chosen from the normal distribution based on a desired confidence level.}

Our technique takes the lower bound of a one-sided confidence interval (the null hypothesis being: $\rho \leq 0$). Thus we obtain a quantile correlation estimate. The quantile $q$ is related to the confidence level of the interval, $\alpha$, by $q=1-\alpha$. Thus an 80\% confidence level yields a $20^{th}$ quantile estimate. We denote this quantile-based estimate as $r_q$,

\begin{equation} \label{eq:quantile-based-cor}
    r_q=\text{tanh}[\text{arctanh}(r - z_{1-q} \times SE_z)]
\end{equation}

\noindent{which is defined so long as four or more matched pairs of observations exist (due to the $\sqrt{n-3}$ term in the denominator of $SE_z$).}

Note: in the statistical programming language \texttt{R}, the quantile estimate is easily obtained with the following code:

\medskip

\begin{lstlisting}[language=R, caption=R example]

    x <- rnorm(100)     # simulate dummy data x, y
    y <- rnorm(100)
    q <- 0.2            # specify desired quantile
    lvl <- (1 - q)      # corresponding conf. level

    # the following returns a quantile estimate of rho
    r.q <- cor.test(x, y, alternative="greater",
                    conf.level=lvl)$conf.int[1]

\end{lstlisting}

\newpage

\section{Simulation code for quantile selection} \label{app:simulation}

The following code can be used to search for optimal quantiles, given a function sim\_dat which simulates bivariate data. The code relies on parallelization, but generates reproducible random numbers for consistency.

\begin{lstlisting}[language=R, caption=quantile-search.R]

library(mvtnorm)
library(parallel)

N.CORES <- detectCores()

# simulation settings
ALPHA <- 0.05
DELTA <- 0
RHO <- c(0.1, 0.25, 0.5, 0.9)
N <- c(20, 50, 100, 200, 500)
PROP.MATCHED <- c(0.1, 0.25, 0.5, 0.75, 0.9)
Q <- seq(0.15, 0.5, 0.05)
N.RUNS <- 10000

# Reproducible random numbers
RNGkind("L'Ecuyer-CMRG")
set.seed(1995)

# simulates a dataset given mean difference, corr., and sample size
sim.dat <- function(mu.diff=0, rho=0, n.pairs=60) {
  rmvnorm(
    n=n.pairs,
    mean=c(0, mu.diff),
    sigma=rbind(c(1, rho), c(rho, 1)),
    checkSymmetry=FALSE)
}

# runs t-test and returns statistic, degrees of freedom, and p-value
t.test2 <- function(x, y, paired=F, rho=0){
  n <- length(x)
  mu.x <- mean(x)
  mu.y <- mean(y)
  d.bar <- mu.x - mu.y
  if (paired) {
    df <- n - 1
    d.all <- x - y
    s2.d <- sum((d.all - d.bar)^2) / (n - 1)
    se <- sqrt(s2.d / n)
  } else {
    df <- 2 * n - 2
    s2.x <- sum((x - mu.x)^2) / (n - 1)
    s2.y <- sum((y - mu.y)^2) / (n - 1)
    s2.p <- (s2.x + s2.y) / 2
    se <- sqrt(s2.p * (2 / n))
    if (rho!=0) {
      se <- se * sqrt(1 - rho)
    }
  }
  p <- 2 * pt(abs(d.bar / se), df=df, lower.tail=F)
  list("t.stat"=(d.bar / se), "df"=df, "p.value"=p)
}

# compute correlation quantile
quantile.cor <- function(x, y, q) {
  cor.test(x, y, alternative="g", conf.level=(1 - q))$conf.int[1]
}

# initialize container for results
quantiles <- array(
    NA,
    dim=c(length(N), length(PROP.MATCHED),
          length(RHO), length(Q)),
    dimnames=list("N"=N, "Prop.matched"=PROP.MATCHED,
                  "Rho"=RHO, "Quantile"=Q))

# `mcapply` splits a job across N cores
for (size in N) {
  for (rho in RHO){
    r <- mclapply(1:N.RUNS, function(i) {
      sim.dat(mu.diff=DELTA, rho=rho, n.pairs=size)
    }, mc.cores=N.CORES)
    
    for (prop in PROP.MATCHED){
      # determine number of matched pairs
      n.matched <- floor(prop * size)
      
      # compute Type I error probability for each quantile
      for (q in Q){
        pvals.q <- mclapply(r, function(dat) {
          if (n.matched > 3){
            r <- quantile.cor(
                dat[1:n.matched, 1],
                dat[1:n.matched, 2],
                q=q)
            t.test2(dat[, 1], dat[, 2], rho=r)$p.value
          } else{
            NA
          }}, mc.cores=N.CORES)
        
        # store type I error probability in container
        quantiles[which(size==N),
                  which(prop==PROP.MATCHED),
                  which(rho==RHO),
                  which(q==Q)] <- mean(pvals.q < 0.05)
      }
    }
  }
}

# determine optimal quantiles based on closest-to-nominal T1 error
quantiles.opt <- apply(
    quantiles,
    MARGIN=1:3,
    function(x) Q[which.min(abs(x - ALPHA))])
quantiles.opt[is.na(quantiles.opt>0)] <- NA

\end{lstlisting}

\end{document}